\documentclass[conference]{IEEEtran}

\usepackage{hyperref}
\hypersetup{
     colorlinks = true,
     citecolor = green,
     }

\ifCLASSINFOpdf
  \usepackage[pdftex]{graphicx}
\else
\fi

\usepackage{framed}
\usepackage{balance}
\balance

\begin{document}
%
\title{Does Co-Development with AI Assistants Lead to More Maintainable Code? A Registered Report}


%
\author{\IEEEauthorblockN{Markus Borg\IEEEauthorrefmark{1}, Dave Hewett\IEEEauthorrefmark{2}, Donald Graham\IEEEauthorrefmark{2}, Noric Couderc\IEEEauthorrefmark{3}, Emma Söderberg\IEEEauthorrefmark{3}, Luke Church\IEEEauthorrefmark{3}, Dave Farley\IEEEauthorrefmark{4}}
\IEEEauthorblockA{\IEEEauthorrefmark{1}CodeScene, Malmö, Sweden and Lund University, Lund, Sweden, markus.borg@codescene.com}
\IEEEauthorblockA{\IEEEauthorrefmark{2}Equal Experts, UK, \{dave.hewett,donald.graham\}@equalexperts.com}
\IEEEauthorblockA{\IEEEauthorrefmark{3}Dept. of Computer Science, Lund University, Lund, Sweden, \{noric.couderc,emma.soderberg\}@cs.lth.se, luke@church.name}
\IEEEauthorblockA{\IEEEauthorrefmark{4}Continuous Delivery, UK, info@continuous-delivery.co.uk}
}


\maketitle

\begin{abstract}
[Background/Context] AI assistants like GitHub Copilot are transforming software engineering; several studies have highlighted productivity improvements. However, their impact on code quality, particularly in terms of maintainability, requires further investigation.
[Objective/Aim] This study aims to examine the influence of AI assistants on software maintainability, specifically assessing how these tools affect the ability of developers to evolve code.
[Method] We will conduct a two-phased controlled experiment involving professional developers. In Phase 1, developers will add a new feature to a Java project, with or without the aid of an AI assistant. Phase 2, a randomized controlled trial, will involve a different set of developers evolving random Phase 1 projects - working without AI assistants. We will employ Bayesian analysis to evaluate differences in completion time, perceived productivity, code quality, and test coverage.
\end{abstract}


%

\section{Introduction}
Using generative AI for code synthesis has become a standard practice among software developers. The pioneering tool, GitHub (GH) Copilot, offers code completion directly within the IDE. Alternatively, developers can use conversational agents such as ChatGPT and Google Bard for code generation. Both approaches are popular according to the JetBrains Developer Survey 2023~\cite{jetbrains_state_2023}, i.e., 77\% of developers report using ChatGPT, and 46\% use GH Copilot.

This study specifically focuses on large language model-based code completion tools within the IDE. We refer to them as AI assistants in this paper. While GH Copilot is the most common tool, alternatives include Amazon CodeWhisperer, JetBrains AI Assist, Visual Studio IntelliCode, and Tabnine. This type of tool represents a new software engineering phenomenon. Early empirical studies show their capabilities in solving relevant tasks~\cite{moradi_dakhel_github_2023}, with some indicating a substantial increase in developer productivity~\cite{peng_impact_2023,ziegler_measuring_2024}.

Ani \textit{et al.} recently published a systematic literature review of GH Copilot studies~\cite{ani_recent_2024}. Several studies address the effectiveness of AI assistants and the productivity gains they offer. Others focus on different aspects of code quality, e.g., security~\cite{perry_users_2023,asare_is_2023} and understandability~\cite{nguyen_empirical_2022,al_madi_how_2023}. We complement previous work through a focus on maintainability. Our starting point is that maintainable code is easy to reason about and change by someone else than the original author. Accordingly, we primarily gauge maintainability by the ease with which a new developer can integrate new features into existing code. Secondly, we enhance our evaluation with additional quality measures to provide a more comprehensive assessment.

\begin{figure}[h]
    \centering
\includegraphics[width=1\linewidth]{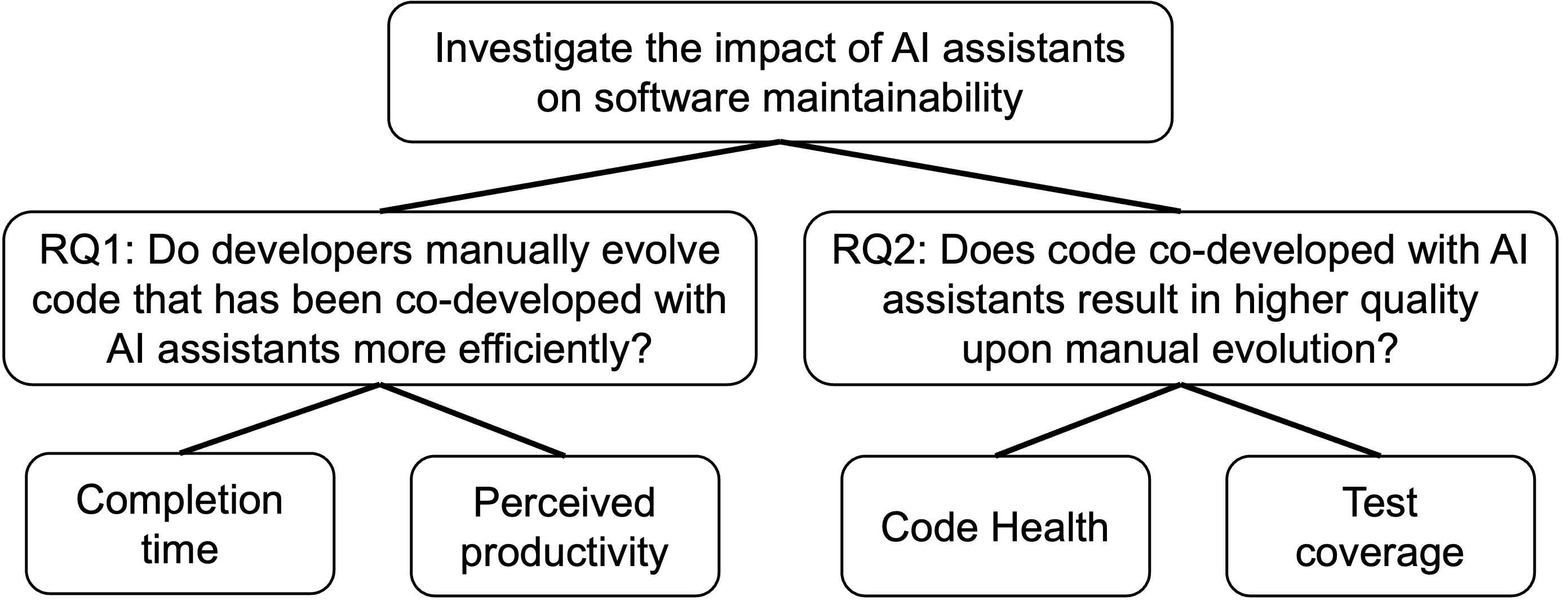}
    \caption{Goal of the study, outlined using the GQM structure.}
    \label{fig:gqm}
\end{figure}

Fig.~\ref{fig:gqm} illustrates the aim of this study, structured using the GQM model~\cite{basili_goal_1994}. Our goal is to investigate the impact of AI assistants on software maintainability. AI assistants will remain part of the development context, thus, we need to better understand how this modified way of working will influence maintenance and evolution.

We break down the goal into two research questions: RQ1) Do developers manually evolve code that has been co-developed with AI assistants more efficiently? and RQ2) Does code co-developed with AI assistants result in higher quality upon manual evolution? We rely on four metrics to answer these RQs: task completion time, perceived productivity according to the SPACE framework, CodeScene's Code Health metric, and test coverage.

We propose a two-phase controlled experiment. In Phase 1, subjects extend a system with or without an AI assistant. In Phase 2, the Randomized Control Trial (RCT), subjects are randomly assigned to evolve a solution from Phase 1 without using AI assistants.

\section{Related work} \label{sec:rw}
About 20 years ago, Simula Research Laboratories conducted a series of controlled experiments on maintainability~\cite{arisholm_series_2010}. The experiments pushed the boundaries of software engineering experiments in terms of the degree of realism, scale, and rigor of experiment conduct. Most strikingly, Simula has pioneered large-scale experiments with professional developers~\cite{benestad_how_2005}. Inspired by Simula's seminal experimental research, we now seek to revisit maintainability in light of AI assistants.

GH researchers have published results from their own studies, but not yet in top academic venues. The most cited work is a controlled experiment published on arXiv~\cite{peng_impact_2023}. Peng~\textit{et al.} recruited 95 freelance programmers and gave them the task of implementing an HTTP server in JavaScript. The participants were randomly assigned to either work with GH Copilot (treatment) or not (control). The study found statistically significant differences in completion time, and the treated group was able to complete the task 55.8\% faster than the control group. There was no significant difference in success rate, i.e., task completion. Our study is similar but focuses on code evolution.

More recently, GH published an experience report in Communications of the ACM~\cite{ziegler_measuring_2024}. Based on a survey focusing on developers' perceived productivity, guided by the SPACE framework~\cite{forsgren_space_2021}, they report that GH Copilot has a large positive impact. The results show that junior developers experience the largest gains. Moreover, the survey suggests that the AI assistant improves a wide range of measures, e.g., task completion time, quality, cognitive load, enjoyment, and learning. We will adapt the SPACE-guided questions used in GH's study, and ask our participants in exit surveys.

Al~Hadi studied the readability of Copilot's generated code in a controlled experiment with students (n=21)~\cite{al_madi_how_2023}. The results suggest that code written by a human pairing with an AI assistant is comparable in complexity and readability to code written by human pair programmers. Our planned study goes beyond Al~Hadi's work by our more realistic setup and involvement of a larger participant pool.

Some studies have highlighted risks with AI assistants in coding tasks. In a user study (n=47), Perry \textit{et al.} found that participants with access to an AI assistant produce less secure code~\cite{perry_users_2023}. Moreover, the AI-assisted participants were also more likely to believe they wrote secure code. Moradi Dakhel compared GH Copilot's code generation with those of students~\cite{moradi_dakhel_github_2023}. While the AI assistant was capable of generating correct solutions for textbook coding tasks, the quality of the code varied -- and some solutions contained defects. Our study aims to provide a balanced perspective on both the opportunities and risks of employing AI assistants in software development. By `risk' we refer to code that might be more efficient to produce yet could be of lower quality and, as a result, more costly to maintain in the long run.

\section{Overview} \label{sec:overview}
\begin{figure}
    \centering
    \includegraphics[width=1\linewidth]{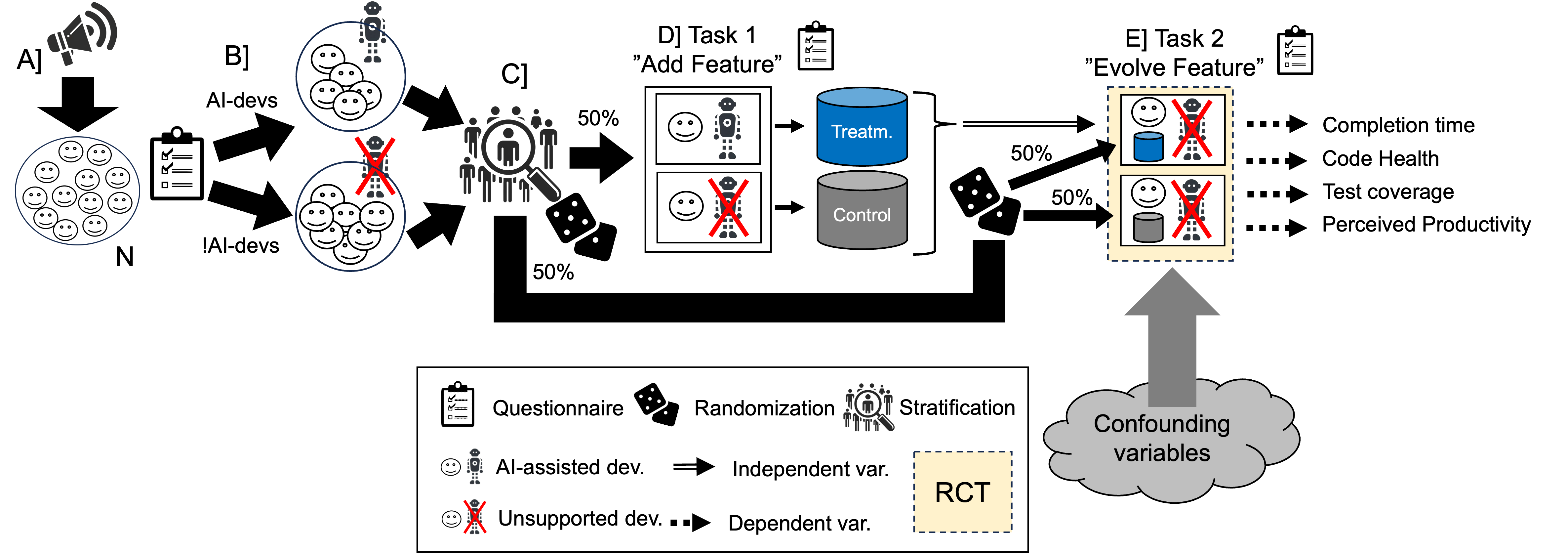}
    \caption{Overview of the study. The part in the yellow box, to which about 50\% of our participants will be assigned, constitutes the RCT.}
    \label{fig:overview}
\end{figure}

Figure~\ref{fig:overview} shows an overview of our two-phase sequential design. In Phase 1, about half of the participants prepare artifacts to be used by either the treatment group or the control group. In Phase 2, the remaining participants take part in an RCT, in which all participants receive only one treatment. Our design is guided by the parts of the ACM SigSoft Empirical Standard for Experiments with Human Participants that are applicable at this stage~\cite{acm_sigsoft_experiments_2024}.

We will call for volunteers to take part in a controlled experiment about ``Software Development with AI Assistants'' (see A] in Fig.~\ref{fig:overview}). The participants will complete assigned tasks remotely in the preferred development environment. Second, as part of the pre-screening of the participants (see Table~\ref{tab:pre-survey}), we ask for i) experience with AI assistants and ii) whether such assistance is currently their preferred way of working (measured using a Likert scale as described in Section~\ref{sec:part}). Participants who answer confirmatory to i) and ii) are split into a separate cohort (see B] in Fig.~\ref{fig:overview}). We refer to the cohorts as \textbf{AI-devs} and \textbf{!AI-devs}, respectively. 

\begin{table*}[h]
\centering
\caption{Outline of the pre-screening questionnaire. A colon indicates a free-text input field. A letter within parentheses shows a mapping to the corresponding SPACE dimension --- item h) is inverted.}
\label{tab:pre-survey}
\begin{tabular}{|p{4cm}|p{1.2cm}|p{12cm}|}
\hline
\textbf{Question} & \textbf{Type/Scale} & \textbf{Operationalization} \\ \hline
Q1-1. Which is your gender? & Nominal & a) Man b) Woman c) Non-binary d) Prefer not to disclose, e) Prefer to self-describe: \\
\hline
Q1-2. What is your age? & Ordinal & a) 19 or younger b) 20-29, ... g) 70 or older \\
\hline
Q1-3. Where do you live? & Nominal & Closed country list \\
\hline
Q1-4. Which of the following best describes what you do? & Nominal & a) Student, full-time or part-time, b) Professional programmer, writing code for work, c) Hobbyist programmer, writing code for fun or outside of work, d) Researcher, e) Other: \\
\hline
Q1-5. Which best describes your programming experience? & Ordinal & a) I’m a student/learning to program, b) 0 to 2 years professional programming experience, ... , f) 
    More than 16 years professional programming experience) \\
\hline
Q1-6. How proficient are you in software development with Java? & Ordinal & a) Beginner, I can write a correct implementation for a simple function b) Intermediate, I can design and implement whole programs, c) Advanced, I can design and implement a complex system architecture \\
\hline
Q1-7. Do you have experience of working with an AI assistant while programming? & Binary & a) Yes, b) No (ends the questionnaire)\\
\hline
Q1-8. Experience and preferences & 5-point Likert + N/A & Thinking of your experience as a developer and your ways of working, please indicate your level of agreement with the following statements.\\
 & & a) I am a habitual user of AI assistants while programming.\\
& & b) I am more productive when using AI assistants. (E)\\
& & c) I complete tasks faster when using AI assistants. (E)\\
& & d) I spend less time searching for information or examples when using AI assistants. (C)\\
& & e) I complete repetitive programming tasks faster when using AI assistants. (E)\\
& & f) Using AI assistants helps me stay in the flow. (E)\\
& & g) Using AI assistants is distracting. (E)\\
& & h) I feel more fulfilled with my job when using AI assistants. (S)\\
& & i) I can focus on more satisfying work when using AI assistants. (S)\\
\hline
\end{tabular}
\end{table*}

Subsequently, we use random stratified sampling to split the participants into either Task 1 or Task 2 of the experiment. Stratification is needed to ensure that we assign an equal share of \textbf{AI-devs} and \textbf{!AI-devs} to Task 1. The recruitment, pre-screening, and stratified sampling are further described in Sec.~\ref{sec:part}. 

In Task 1, the \textbf{AI-devs} and \textbf{!AI-devs} cohorts add a new feature to an existing Java system (see D] in Fig.~\ref{fig:overview}). Later, the code that \textbf{AI-devs} and \textbf{!AI-devs} submit will be provided to another developer in Task 2 for further evolution. Task 1 will be concluded by the exit questionnaire described in Table~\ref{tab:exit-survey} and the code quality will be measured.

\begin{table*}[h]
\centering
\caption{Tentative exit questionnaire. A colon indicates a free-text input field. A letter within parentheses shows a mapping to the corresponding SPACE dimension --- items d) and j) are inverted.}
\label{tab:exit-survey}
\begin{tabular}{|p{6cm}|p{1.2cm}|p{10cm}|}
\hline
\textbf{Question} & \textbf{Type/Scale} & \textbf{Operationalization} \\ \hline
Q2-1. Did you complete the task in one uninterrupted sitting? & Nominal & a) Yes, b) Yes, but with short breaks, c) No\\ \hline
Q2-2. Did you use AI assistants during the development of the task? & Binary & a) Yes,  c) No\\
\hline
Q2-3. (AI-devs only) Which AI assistant did you work with? & Closed list & a) GitHub Copilot, b) AWS CodeWhisperer, c) JetBrains AI Assistant, d) Visual Studio IntelliCode, e) TabNine, f) Other: \\\hline
Q2-4. (AI-devs only) How frequently did you interact with the AI assistant? & Ordinal & a) Hardly at all, b) Sometimes, c) Often, d) Almost for every statement I wrote\\
\hline
Q2-5. Please list any development tools used during the task beyond the standard IDE. Examples include code linting tools, quality analyzers, and vulnerability scanners. & Nominal & a) N/A, b) Used tools: \\
\hline
Q2-6. Perceived productivity & 5-point Likert + N/A & Thinking of your experience with the task, please indicate your level of agreement with the following statements.\\
 & & a) I was focused on the task during the programming session. (E)\\
  & & b) I was a productive programmer while completing the task. (E)\\
& & c) I felt fulfilled while completing the programming task. (S)\\
& & d) I found myself frustrated while completing the programming task. (S)\\
& & e) I made fast progress despite working with an unfamiliar system. (E)\\
& & f) The code I wrote was of high quality. (S)\\
& & g) I maintained a state of flow during the programming task. (E)\\
& & h) I enjoyed completing this task. (S)\\
& & i) I completed the repetitive programming activities fast during the task. (E)\\
& & j) I spent considerable time searching for information or examples during the task. (C)\\
    & & k) The task generally resembled development work I have done in the past. \\
\hline
Q2-7. Is there anything you would like to add regarding the study or your role in the experiment? & Free-text & : \\
\hline
Q2-8. Please provide your email if you want to receive the report when the study is done. & Free-text & : \\
\hline
\end{tabular}
\end{table*}


For Task 2, which constitutes the RCT, new participants are randomly assigned (without replacement) to evolve a valid Task 1 solution (see E] in Fig.~\ref{fig:overview}). The solutions are either from the \textbf{AI-dev} cohort (the treatment) or the \textbf{!AI-dev} cohort (the control). Note that Task 2 will be completed without AI assistants. Both tasks are further described in Sec~\ref{sec:task}. Task 2 will use the same exit questionnaire as Task 1.

\section{Variables} \label{sec:var}
We propose a two-level single-factor experiment with underlying variations within the two levels. The \textbf{independent variable} is whether the participants evolve code that has been co-developed with an AI assistant during Phase~1 (the treatment group) or not (the control group). We consider this as a binary variable in the primary analysis but will control for the Phase~1 quality variation in the secondary analysis. As shown in Figure~\ref{fig:overview}, we measure four \textbf{dependent variables} in the Phase 2 RCT.

\begin{itemize}
    \item Completion time: The time between a participant receives the task and submits the final solution. Time is measured in minutes on a ratio scale.
    \item Code Health (CH): The CH of the solution in Phase 2. This is measured as a weighted average of the files in the code base. Previous work shows that CodeScene's CH is associated with defect counts and development effort~\cite{tornhill_code_2022,borg_increasing_2024}. CH is a number between 1 and 10 measured on an interval scale (no true zero).
    \item Test coverage (TC): The statement coverage of the test suite from the final solution in Phase 2. This is measured as a percentage on an interval scale.
    \item Perceived productivity (PP): A subjective assessment as proposed by the SPACE framework~\cite{forsgren_space_2021}. We measure this variable using a Likert scale, composed of ordinal Likert items, in a manner analogous to the method described by Ziegler \textit{et al.}~\cite{ziegler_measuring_2024}. After having checked the internal consistency of the Likert scale using Cronbach's Alpha, we will calculate an average score for the items (since N/A is an option) -- taking into consideration that two items are inverted. 
\end{itemize}

Several \textbf{confounding variables} are inevitably at play in software development experiments. Furthermore, our design with remote participants enhances realism at the expense of control. Although the randomization used in the RCT balances both observed and unobserved confounders across the treatment and control groups, we list the most important confounders in Table~\ref{tab:confounders}. Sec.~\ref{sec:part} explains why we control for developer experience upfront, while other confounders are measured to enable post hoc analysis. 

\begin{figure}
    \centering
    \includegraphics[width=1\linewidth]{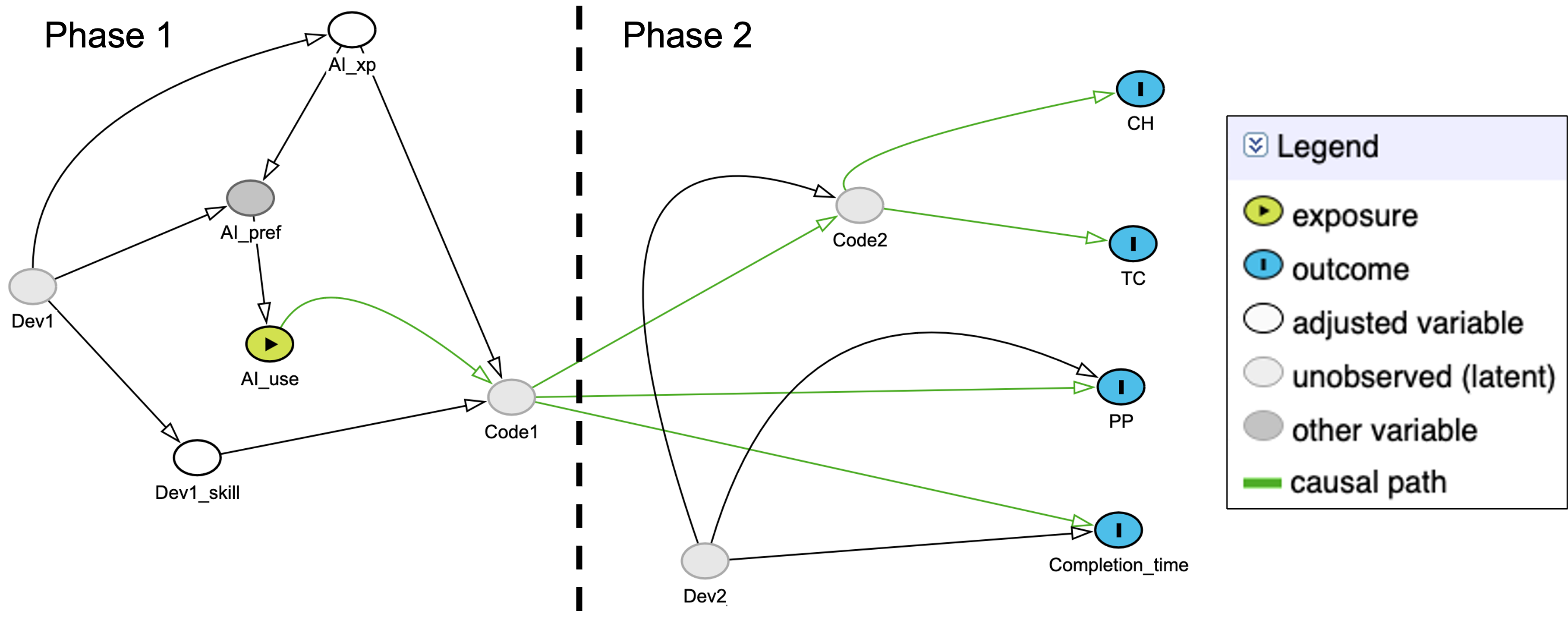}
    \caption{DAGitty causal graph. Dev1 and Dev2 represent the full complexity of the human participants in Phases 1 and 2, respectively. Code1 and Code2 are the participants' solutions after Phases 1 and 2, respectively. $AI\_use$ is the independent variable. The other variables are explained in Table~\ref{tab:confounders}.}
    \label{fig:causal}
\end{figure}

Figure~\ref{fig:causal} shows our proposed \textbf{causal graph} created using DAGitty (further explained on Zenodo~\cite{equal_experts_does_2024}). In the first phase leading to Code1, we state that $AI\_use$ (the independent variable), $AI\_xp$, and $Dev1\_skill$ causally influence Code1. By adjusting for $Dev1\_skill$ and $AI\_xp$, we find a green causal path in DAGitty that continues through Phase 2 to the four dependent variables (CH, TC, PP, and Completion time). 

In the second phase, Code1 and Dev2 causally impact Code2. We measure the causal effects using the two dependent variables CH and TC. The dependent variables PP and Completion time are causally influenced by Code1 and Dev2.

\begin{table*}[]
\label{tab:confounders}
\caption{Main confounding variables in the RCT. The questionnaires are available in Tables~\ref{tab:pre-survey} and~\ref{tab:exit-survey}.}
\begin{tabular}{|p{2.4cm}|p{7.3cm}|p{1cm}|p{5.6cm}|}
\hline
\textbf{Name}                         & \textbf{Description}                                                            & \textbf{Scale}    & \textbf{Operationalization}                               \\ \hline
$Dev1\_skill$ & How skilled the participant is as a (Java) developer  & Ordinal  & Controlled: 6-point scale, questionnaire (Q1-5) (Q1-6 used to filer out Java novices)   \\ \hline
$AI\_xp$      & How used the participant is to work with AI assistants & Ordinal  & Controlled: 5-point scale, questionnaire (Q1-8a) \\ \hline
$AI\_pref$      & Whether the participant prefers to work with AI assistants & Ordinal  & Controlled: 5-point scale, questionnaire (Q1-8) \\ \hline
Uninterrupted work           & Whether the participant completed the task in one sitting              & Ordinal  & Measured: 3-point scale, questionnaire (Q2-1)    \\ \hline
AI tool       & Which AI assistant was used during Phase 1                      & Nominal  & Measured: Closed list, questionnaire (Q2-3)  \\ \hline
AI influence       & How much the AI assistant was used during Phase 1                      & Ordinal  & Measured: 4-point scale, questionnaire (Q2-2,4)  \\ \hline
Tool usage              & Whether the participant used supportive development tools                             & Nominal & Measured: Free-text, questionnaire (Q2-5)                \\ \hline
Phase 1 quality              & The code quality when starting Phase 2                                 & Interval & Measured: CodeScene's CH              \\ \hline
\end{tabular}
\end{table*}


\section{Hypotheses} \label{sec:hypo}
We hypothesize that the use of AI assistants will have a positive impact on the dependent variables in the Phase 2 RCT. We propose a hybrid approach with Bayesian analysis complemented with conventional hypothesis testing. On the one hand, the Bayesian analysis will explore uncertainties to provide a robust probabilistic understanding. On the other hand, the results from the frequentist hypothesis testing will facilitate communication of results to industrial practitioners.

From the frequentist perspective, our study investigates four null hypotheses ($H_{0}1$ -- $H_{0}4$) corresponding to the dependent variables (see Fig.~\ref{fig:overview}). We formally state that in the RCT, there is \textit{no difference} in the:   

\begin{itemize}
    \item[$H_{0}1$] Completion time
    \item[$H_{0}2$] CH
    \item[$H_{0}3$] TC
    \item[$H_{0}4$] PP
\end{itemize}

\noindent between participants evolving code by \textbf{AI-devs} (treatment) and \textbf{!AI-devs} (control). For each null hypothesis, we specify a corresponding non-directional alternative hypothesis ($H_{A}1$ -- $H_{A}4$), stating that there \textit{is a difference}.

We expect the dependent variables Completion Time, CH, and TC to be normally distributed. After conducting normality tests, we plan to conduct t-tests using the conventional significance level $\alpha = 0.05$. We will complement the p-values by reporting effect sizes. PP, on the other hand, is measured using a Likert scale. Differences between groups will be tested using a rank-sum test, which is suitable for ordinal data.

\section{Participants} \label{sec:part}
All participants are volunteers who choose to engage in the research. We will recruit participants through i) social media advertisements on platforms such as YouTube, LinkedIn, and X and ii) using our personal networks. Although our target population consists of professional software developers, we will not restrict participation. The pre-screening questionnaire outlined in Table~\ref{tab:pre-survey} will allow us to filter participants as necessary.

The pre-screening questionnaire also facilitates random stratified sampling to ensure an equal distribution of participants across the \textbf{AI-dev} and \textbf{!AI-dev} cohorts for Task 1. To qualify for the \textbf{AI-dev} cohort, participants must: i) answer yes to Q1-7, ii) agree to statement Q1-8a), and iii) have a positive mean response to the preference questions Q1-8 b)--i) --- taking the inverted question g) into account. Allowing participants to adhere to their preferred work methods reduces the likelihood of non-compliance, such as using AI tools when told not to. However, this flexibility might introduce a bias, i.e., less experienced developers could be more inclined to use AI assistants -- a benefit suggested by early research~\cite{ziegler_measuring_2024}. To compensate for this expected bias, we will control for developer experience in our analysis.

We used G*Power (v.3.1.9.7) to run a power analysis for a two-tailed t-test to set a target for participant recruitment. Based on our expectation of medium effect sizes (d=0.5) for the dependent variables, $\alpha = 0.05$, and a power of 0.80, the result indicates that we need at least 64 participants per group for the Phase 2 RCT to allow frequentist hypothesis testing. Therefore, as we seek equally many participants for Phase 1, the target number of participants is 256 -- a high number in software engineering research. However, with our extensive social media reach and the contentious nature of the AI assistant topic, which will be transparently communicated, we expect to reach this target.

The design of our study adheres to the essential attributes of the ACM SigSoft Empirical Standard ``Ethics (Studies with Human Participants)''~\cite{acm_sigsoft_ethics_2024}. Before assigning tasks to participants, we will ensure they understand their privacy is protected, their data will be handled in compliance with GDPR, and that they may withdraw from the study at any time without consequence. We will explain our intention to share our findings through academic publications, as well as blog posts and white papers to reach the general software development community. After these disclosures, we will obtain informed consent from all participants.

There are no anticipated risks of harm to participants. However, they will need to invest 2--4 hours to complete the assigned task. This is similar in nature and duration to programming tests used in recruitment processes. We believe this is reasonable and the task offers a learning opportunity without mundane or valueless activities. Participants will be incentivized to complete the task by the chance to win either a grand prize (about 100\$) or one of 100 books signed by the last author. Finally, submitting this registered report for peer review acts as an independent assessment that the study design meets the ethical standards of the empirical software engineering community. 

\section{Task and Dataset} \label{sec:task}
We have created two tasks that embody contemporary Java development. The task evolved over several weeks with input from developers at Equal Experts. At a brainstorming workshop with four senior consultants and the first author, we defined the design goal as specifying a standard, real-world problem that should be recognizable by any professional developer. This approach helps mitigate threats to internal validity related to task familiarity. The problem includes the following attributes: 
\begin{itemize}
    \item code spread across multiple files, but understandable.
    \item unit tests present, but not complete coverage.
    \item API and database integration.
    \item involving a well-known framework.
    \item possible to complete in 2--4 hours.
    \item problem statement should be fun/interesting.
    \item easily understandable domain.
    \item an injected bug in the code.
\end{itemize}

Guided by the above, participants will be introduced to a hypothetical new business, Recipes4Success (R4S), whose mission is to ignite a passion for cooking among young people. Participants learn that R4S had previously engaged a software consultancy to develop a recipe service. Unfortunately, the collaboration did not meet expectations --  a working web app was eventually delivered but with poor software quality. R4S now seeks to enhance this service with new features and to establish a fruitful partnership with another consultancy, setting the stage for the participant's involvement.

The total size of the code base sent to the participants is 2.5 KLoC across 47 files. The base application, recipe-finder, is a deliberately substandard Java/Spring Boot application used as the starting point for Task 1. A specific bug relevant to the task has been intentionally introduced. The base application supports two functionalities: i) filtering the recipe list by a search term, and ii) displaying all recipes when no search term is provided.

The code base contains two test suites. First, a suite of unit tests helps participants understand the details of Java methods. Moreover, the test suite provides a starting point for participants who prefer working with test-driven development. Participants can then add or modify the unit tests as they add new features. Note that the unit tests do not expose the injected bug. Second, the acceptance test suite targets the expected behavior of the completed solution. The participants must resolve the injected bug to make the acceptance tests pass. Participants will be instructed not to modify the acceptance tests, and we will monitor compliance.


Participants will complete one of two tasks. In Task~1, they are tasked with enhancing the existing search feature to include filtering recipes by the total time required to prepare a meal. The participants are specifically instructed to ensure that their contributions are of high quality and maintainable. Task~2 introduces new participants to the same R4S scenario as described above. They are asked to further develop the Task 1 search feature by incorporating a filter for the cost per serving, requiring them to evolve the code that was initially developed with or without AI assistants.

We have iteratively developed the code base and the tasks based on feedback from pilot tests. Five consultants from Equal Experts have tested either Task~1 or Task~2, ensuring that the instructions are clear, the estimated time budget is reasonable, and that the infrastructure works (see Sec.~\ref{sec:exec}).

To maintain the integrity of the study, i.e., preventing any leakage to AI assistants' training data~\cite{silva_gitbug-java_2024}, we choose not to host the code in a public git repository. Instead, all relevant documents and code are shared through a replication package on Zenodo, available as PDF documents~\cite{equal_experts_does_2024}.

Clearly, the conclusions drawn from studying this specific task cannot be generalized to all possible development scenarios. Nonetheless, it remains a highly relevant task to study as it resembles a typical industrial assignment,  progressing work previously completed by an unknown developer. The main threats to external validity stem from our choice of a single programming language and the complexity of the task. We believe this code base is the largest feasible size for participants to understand and extend within a reasonable time.

\section{Execution Plan} \label{sec:exec}
\textbf{Task distribution.} After participants voluntarily sign up, they will be asked to complete a pre-screening questionnaire. Based on their responses, we will distribute them into either one of the Phase 1 cohorts or the Phase 2 RCT. Both programming tasks will be administered using the \textit{snapcode.review}\footnote{\url{https://snapcode.review}} platform, which is used by Equal Experts for coding tests during recruitment processes. snapcode.review is a service that automates take-home challenges and evaluations using GitHub. This solution enables us to trigger measurements and collect time stamps in a simple and scalable way. 

Completion time will be measured from 1) the moment a participant gains access to a remote GitHub repository, which contains their unique copy of the code base, to 2) the point at which they irrevocably click a submit button to conclude the task. Between 1) and 2), participants are free to push commits to the GitHub repository as frequently as they choose. This flexibility allows each participant to employ their preferred git workflow.

\textbf{Data cleaning.} For both tasks, we will exclude participants whose solutions fail to pass the acceptance test suite. Additionally, we will remove participants who indicate in the exit questionnaire that they did not adhere to the task instructions, ensuring data integrity and adherence to the study protocol. Moreover, we will remove outlier submissions that indicate any attempts by participants to game or manipulate the system.

\textbf{Data analysis.} 
We plan to utilize a mixed-methods approach in our analysis. For all data analysis, we assume independence of observations, i.e., that the participants do not communicate. We will begin with a Bayesian analysis to gain a probabilistic understanding of the treatment effects, an approach that remains robust even in the case of smaller sample sizes. Specifically, we will construct three types of priors: i) uninformative priors, ii) AI-sceptical priors, and iii) AI-enthusiastic priors, to reflect different stances on the maintainability impact of AI assistance --- this also serves as a sensitivity analysis. Further details are provided online~\cite{equal_experts_does_2024}.

Following the Bayesian analysis, we will perform statistical hypothesis testing as detailed in Sec.~\ref{sec:hypo}. While it can be observed as redundant, we believe it will facilitate the dissemination of results in a format that practitioners find more accessible. In the primary analysis, we will treat the independent variable as binary. If we reach the targeted numbers, we expect the dependent variables to be normally distributed (which will also shape our Bayesian priors). We will perform standard t-tests and report effect sizes and confidence intervals in accordance with the empirical standard.

In the secondary analysis, we will control for the main confounding variables listed in Table~\ref{tab:confounders} and potential interaction effects. Using the ordinal measurements, we can adjust for the influence of five potential confounders. Furthermore, we believe that the use of various (non-AI) supportive tools in the IDE can largely impact the dependent variables. We rely on the participants' self-reporting of such tools and will analyze the results for major confounding effects. Finally, we will assess the quality of the Task 1 solutions. Should substantial variations be observed within the two cohorts, we reserve the option to employ blocking in our statistical analysis. 

\textbf{Time Schedule}. Recruitment is scheduled for a two-week period in late summer 2024. If sufficient numbers are achieved swiftly, Task 1 will be distributed with a subsequent two-week completion window. Following this, Task 2 will be distributed, also with a two-week completion deadline. The schedule takes advantage of the potentially lighter workloads during the (northern hemisphere) summer months in many organizations. In case recruitment is slow, we will use a rolling window approach to introduce new participants in batches.

\section{Risk Analysis and Threats to Validity} \label{sec:risk}
The main risk in this study is failing to attract enough participants. We are well aware of this and have carefully planned how to advertise the study.

A related risk is the potential imbalance in the distribution of \textbf{AI-devs} and \textbf{!AI-devs} among the volunteers. If most volunteers prefer working with AI assistants, we must request some to disable tools for this study. This might make participants unhappy, possibly leading to dropouts or biased results.

The construct of an ``AI assistant'' is central to our study. As AI is notoriously difficult to define, there is a risk that our participants will interpret the construct differently. This might lead to incorrect responses on questionnaires. To mitigate this, we will provide a precise definition in the instructions, with examples, and repeat them in the questionnaires.

Despite clear instructions, there remains a risk that some participants may not adhere to them. For example, using ChatGPT despite being told not to. Our choice to conduct the study in remote, realistic development environments rather than controlled classroom settings means that we cannot directly monitor compliance.  We do not want to track activities on remote machines, opting to rely on participants' self-reporting in the exit questionnaires.

\bibliographystyle{IEEEtran}
\bibliography{IEEEabrv, llm-prog}

\end{document}